\documentclass[letterpaper, 10 pt, conference]{ieeeconf}  

\IEEEoverridecommandlockouts                              

\usepackage{graphics} 
\usepackage{epsfig} 
\usepackage{mathptmx} 
\usepackage{times} 
\usepackage{amsmath} 
\usepackage{amssymb}  
\usepackage{cite}
\usepackage{cuted}
\usepackage{amsmath,amssymb,amsfonts}
\usepackage{algorithmic}
\usepackage{graphicx}
\usepackage{textcomp}
\usepackage{xcolor}
\usepackage{bm}
\usepackage{hyperref}

\title{\LARGE \bf
Innovative Gain Reconfiguration for Active Fault-Tolerant Flight Control: Balance of Stability and Agility}

\author{Ege C. ALTUNKAYA,
        Akin CATAK,
        Emre KOYUNCU,~\IEEEmembership{Senior Member,~IEEE,}
        and Ibrahim OZKOL
\thanks{The authors are with the Aerospace Research Center, Istanbul Technical University, Istanbul, 34469, Türkiye (e-mail: altunkaya16@itu.edu.tr; catak15@itu.edu.tr; emre.koyuncu@itu.edu.tr; ozkol@itu.edu.tr)}
\thanks{(Corresponding author: Ege C. ALTUNKAYA (e-mail: altunkaya16@itu.edu.tr))}}

\usepackage{longtable,tabularx}

\usepackage{comment}
\usepackage[super]{nth}

\newtheorem{assumption}{Assumption}[section]
\newtheorem{corollary}{Corollary}[section]

\usepackage{hyperref}
\usepackage{lipsum}

\begin{document}

\maketitle

\begin{abstract}
In this study, a distinct reconfigurable fault-tolerant flight control strategy is addressed for mitigating one of the persistent safety-critical issue, i.e. loss of control triggered by actuator faults. The attainable acceleration set notion is taken a step further towards incremental attainable acceleration set through a slight modification that enables instantaneous controllability checks. The inner-loop gains are updated in case of a fault using incremental attainable acceleration set and a tuning function, which is in charge as a compensator of agility and robustness. Additionally, the outer-loop gains are also such reconfigured that holding the bandwidth ratio of the successive loops at a prudent level to ensure the closed-loop stability; for this reason, an analytical outer-loop gain update law is derived based on the inner-loop gains and bandwidth, actuator and command filter time constants. Subsequently, the proposed architecture is assessed under a severe fault scenario with a demanding maneuver mission. Noticeably, the proposed method fulfills the expectations of stability and agility sufficiently, and surpasses the fixed-gain approach.
\end{abstract}


\section{Introduction}
\label{intro}

Ensuring the safety of the aircraft under any circumstance is the main objective of the entire design process, that's why the civilian or military certifications provide worthy guidelines to track during the design \cite{mil-f, CSVLA}. However, unfortunately, regardless of how safely the aircraft is designed, numerous fault-prone elements are mounted on an aircraft. Therefore, as one of the safety assessments, the question of the likelihood of a fault occurrence per flight can be partially answered at the system or whole aircraft level. Since the impact of a fault can be dependent on the flight conditions, estimating the potential effects on the aircraft is a tedious workload. At this point, the necessity of an adaptiveness of the aircraft against any type of fault under any circumstance arises.

Nevertheless, one of the disciplines which necessitates an adaptiveness against presumable faults is the flight control, especially for an unstable fighter aircraft. There exist numerous engineering and theoretical solutions for accommodating these faults, such as robust flight control design or active fault-tolerant control design \cite{fekihSurvey, surveyExtra, surveyExtra2}. Each technique, inherently, comes with advantages, disadvantages, and even prerequisites, e.g. over-actuation requires a control allocation method while robust flight control design may compromise the agility of the aircraft \cite{fekihSurvey}. However, the active fault-tolerant control has attracted attention for a while instead of robust flight control design for fault-tolerance \cite{surveyExtra}.

In the open literature, there are plenty of active fault-tolerant flight control proposals \cite{lovely3, ergocmen1, ergocmen2, mallavalli, tohidi, vile, kufoalor, peni}. Recently, the attention of researchers has been drawn to reinforcement learning and artificial neural network-based fault accommodation \cite{cdc1, cdc2}. Furthermore, control allocation-based fault-tolerant control design methods continue to be proposed, such as \cite{cdc3, su2023}; however, it should not be confused that the existence of redundancy in the control effectors implies that the system can cope with presumable actuator faults under diverse circumstances. Therefore, care must be taken even if the system is designed in an over-actuated suite. Some researchers recognize the severity of the situation where an actuator fault during maneuvers could trigger instability, emphasizing that fault-tolerant control design cannot solely rely on handling model uncertainties \cite{wang2023}.

Additionally, in \cite{ergocmen1}, the author proposes dynamic control allocation depending on actuator fault knowledge to accommodate various types of actuator faults, such as loss of effectiveness, lock-in-place, and control surface damage. However, the necessity of extensive and case-specific planning exists to reconfigure the control allocation architecture. Furthermore, in \cite{ergocmen2}, the authors propose an active fault-tolerant flight control scheme by limiting the achievable kinematics of the aircraft during a maneuver to prevent a loss of control due to actuator faults, which is successfully simulated. Again, however, offline and extensive planning is required to accommodate faults in-flight. Vile et al. \cite{vile} propose a priority-weighted control allocation scheme and sliding mode controller. While the proposed method performs satisfactorily, it relies on offline prioritization, which is a conventional approach. Also, since the flight control law is derived by leveraging sliding mode control, passive fault tolerance, i.e., robust control characteristics, may dominate the response instead of prioritization.

\subsection{Contributions}
\label{contr}

The main contributions of this study are as follows;

\begin{itemize}
    \item This study presents a unique perspective on active fault-tolerant flight control by addressing the problem through aerospace notions. In this context, the incremental attainable acceleration set (IAAS) is introduced and coherent collaboration of nonlinear dynamic inversion and IAAS is proposed for fault accommodation.
    
    \item A new gain reconfiguration mechanism is developed that leverages both IAAS and a tuning function which harmonizes the stability and agility of the aircraft, which is one of the fundamental problems of the active fault-tolerant control.

    \item Additionally, an analytical outer-loop gain update law is derived principally based on inner-loop reconfiguration and successive loop bandwidth ratio specification.
\end{itemize}

Consequently, the proposed method's efficacy is examined through a coupled maneuver mission.

\section{Preliminaries}
\label{preliminaries}

The necessary background for the subsequent sections is established in this section.

\subsection{Notations}
\label{notations}

Throughout the study, the time derivative of a $C^1$ function $f : \mathbb{R}^n \xrightarrow[]{} \mathbb{R}$ is denoted by $\dot{f}$. Moreover, a vector is denoted in the bold type, i.e. $\bm{v}$, and the vector product of two vectors $\bm{x}$ and $\bm{y}$ is denoted by $\bm{x} \times \bm{y}$. The operator of $\oslash$ denotes the Hadamard division, and $sgn(*)$ is the signum function. Define $\gamma \in \mathbb{R}$, where $\bm{\gamma}_m$ denotes the row vector of size $m$ with elements given by $\gamma$. Furthermore, define a row vector of $\bm{\alpha} \in \mathbb{R}^p$ and a number of $\rho \in \mathbb{Z}_{>0}$, the notation of $\bm{\alpha}_\rho$ denotes the $\rho^\text{th}$ element of $\bm{\alpha}$. Similarly, if $\mathbb{A} \in \mathbb{R}^{n \times n}$ is a diagonal matrix, the notation of $\mathbb{A}_\rho$ denotes the $\rho^\text{th}$ diagonal element of $\mathbb{A}$. Finally, a control affine system is described as given in equation~\eqref{controlAffineSystem}.

\begin{equation}
\label{controlAffineSystem}
    \bm{\dot{x}} = \bm{f}(\bm{x}) + \bm{g}(\bm{x})\bm{u}
\end{equation}  
where $\bm{x} \in \mathbb{R}^n$ is the state vector, and $\bm{u} \in \mathbb{R}^m$ is the control input vector. Nonlinear mappings of $\bm{f} : \mathbb{R}^n \xrightarrow[]{} \mathbb{R}^n$ and $\bm{g} : \mathbb{R}^n \xrightarrow[]{} \mathbb{R}^{n \times m}$ are locally Lipschitz continuous functions.

\subsection{Flight Dynamics Model}
\label{fdm}

The 6-degrees-of-freedom nonlinear translational and rotational dynamics of an aircraft are given in a compact form in Eq.~\eqref{translationalDynamics} and Eq.~\eqref{rotationalDynamics}. 

\begin{equation}
\label{translationalDynamics}
    \bm{\dot{V}} = m^{-1}[\bm{F} -  \bm{\omega} \times m \bm{V}]
\end{equation}  

\begin{equation}
\label{rotationalDynamics}
    \bm{\dot{\omega}} = J^{-1}[\bm{M} -  \bm{\omega} \times J \bm{\omega}]
\end{equation}  
where $\bm{V} \in \mathbb{R}^{3 \times 1}$ is the body velocity vector, $\bm{\omega} \in \mathbb{R}^{3 \times 1}$ is the angular rate vector, $\bm{F} \in \mathbb{R}^{3 \times 1}$ is the total body force vector, $\bm{M} \in \mathbb{R}^{3 \times 1}$ is the total moment vector, $J \in \mathbb{R}^{3 \times 3}$ is the inertia tensor, and $m$ is the mass of the aircraft. Finally, the rotational kinematics are presented in equation~\eqref{rotationalKinematics}.

\begin{equation}
\label{rotationalKinematics}
    \bm{\dot{\Omega}} = 
\begin{bmatrix}
1 & \sin\phi\tan\theta & \cos\phi\tan\theta \\
0 & \cos\phi & -\sin\phi \\
0 & \sin\phi \sec\theta & \cos\phi \sec\theta
\end{bmatrix} \bm{\omega}
\end{equation}  
where $\bm{\Omega}$ is Euler's angle vector. 

Furthermore, the over-actuated aerodynamic modeling is derived from \cite{aerodynamicPolynomials, forceMomentEquations}, where each coefficient is formulated as a polynomial function of the corresponding states and control surface deflections. The formulation details can be found in \cite{aerodynamicPolynomials}.

\subsection{Fault Detection and Isolation}
\label{fdi}

The fault diagnosis modeling is not within the scope of the study; therefore, it is assumed that such a fault detection and isolation methodology is implemented, which detects and isolates the actuator jam fault within 0.1 seconds.

\subsection{Incremental Attainable Acceleration Set}
\label{iaas}

The attainable acceleration set (AAS) is a distinct formation of the attainable moment set (AMS) by leveraging the Euler's equations of motion. Since these sets give insight into the capabilities and controllable boundaries of the aircraft, benefiting from AAS is reasonable in order to assess whether the demanded commands are still bounded by AAS or not.

However, as a one step further, deriving the AAS by including the actuator rate limits allows to investigate the instantaneous controllability. With this rationale, primarily, the incremental attainable moment set notion is introduced as given in equation~\eqref{eq:iams1}.

\begin{equation}\label{eq:iams1}
        \Delta \mathbb{A_M} = \{\Delta \bm{\tau} | \Delta \bm{\tau} = \Phi\Delta \bm{u}, \hspace{0.15cm} \underline{\Delta \bm{u}} \leq \Delta \bm{u} \leq \overline{\Delta \bm{u}}\}
\end{equation}
where $\Delta \bm{\tau} \in \mathbb{R}^{3 \times 1}$ represents the vertices of the convex-hull in question, while $\Delta \bm{u}$ denotes the control input in the incremental form. Additionally, $\underline{\Delta \bm{u}}$ and $\overline{\Delta \bm{u}}$ denotes the lower and upper limits of the control inputs by considering the actuator constraints as defined in equation~\eqref{eq:iams2}.

\begin{equation} \label{eq:iams2}
    \begin{split}
    \overline{\Delta \bm{u}} &= \textrm{min}(\dot{\bm{u}}_{max} \Delta t_s, \bm{u}_{max} - \bm{u}_0) \\
    \underline{\Delta \bm{u}} &= \textrm{max}(-\dot{\bm{u}}_{max} \Delta t_s, \bm{u}_{min} - \bm{u}_0)
    \end{split}
\end{equation}
in which $\dot{\bm{u}}_{max}$ is the maximum actuator rate limit, whereas $\bm{u}_{max}$ and $\bm{u}_{min}$ are the maximum and minimum actuator position limits, respectively. Also, $t_s$ is the fixed time step magnitude and $u_0$ is the current actuator position. Based on the incremental AMS derivation, the incremental AAS can be expressed as given in equation~\eqref{eq:aas1}.

\begin{equation}
\label{eq:aas1}
    \Delta \mathbb{A_A} =\{\Delta \dot{\bm{\omega}}| \Delta \dot{\bm{\omega}} =J^{-1}[\Delta \bm{\tau} - \bm{\omega} \times J \bm{\omega}], \hspace{0.15cm} \Delta \bm{\tau} \in \Delta \mathbb{A_M}\}
\end{equation}
in which $\bm{\omega} = [p \hspace{0.15cm} q \hspace{0.15cm} r]^T$, representing angular velocities. Consequently, the incremental AAS is derived to understand if the aircraft is capable of performing the desired commands at that instant, taking into account aerodynamic and mechanical characteristics, as well as actuator constraints. 

\begin{assumption}[\textbf{Convex structure of the attainable sets}]
    The convex structure of the attainable acceleration set under healthy circumstances may deteriorate after a fault occurrence. However, throughout the study, it is assumed that the convex form is not significantly affected, thus the convex-hull approach is still reasonable.
\end{assumption}

\section{Flight Control Architecture}
\label{fca}

\subsection{Control law design}
\label{cld}

The control law for the inner loop is designed using nonlinear dynamic inversion. Since the aircraft is over-actuated, the inner-loop control law is such derived that the output of the flight control module is to be the desired control moment coefficient to feed the control allocation. Therefore, using Euler's equations of motion in equation~\eqref{rotationalDynamics}, the necessary moment coefficients can be expressed in the inverse form of equation~\eqref{controlAffineSystem} as given in equation~\eqref{eq:eq12}.

\begin{equation} \label{eq:eq12}
\begin{bmatrix}
C_l \\
C_m \\
C_n
\end{bmatrix}_c
=
\frac{J}{\Bar{q}_\infty S}
\begin{bmatrix}
b & 0 & 0 \\
0 & \Bar{c} & 0 \\
0 & 0 & b
\end{bmatrix}^{-1}
\Big\{\bm{\Dot{\omega}}_c + J^{-1}\Big(\bm{\omega}\times J\bm{\omega}\Big)\Big\}
\end{equation}
in which $b, \Bar{c}, S, \Bar{q}_\infty$ are the wing span, mean aerodynamic chord, wing area, and dynamic pressure, respectively. $[C_l \hspace{0.15cm} C_m \hspace{0.15cm} C_n]_c^{T}$ corresponds to the commanded roll, pitch, and yaw moment coefficients. Also, $\bm{\Dot{\omega}}_c = [\dot{p} \hspace{0.15cm} \dot{q} \hspace{0.15cm} \dot{r}]_c^{T}$ is the virtual input as given in equation~\eqref{eq:eq13}.

\begin{equation} \label{eq:eq13}
\begin{bmatrix}
\Dot{p} \\
\Dot{q} \\
\Dot{r} \\
\end{bmatrix}_c
= 
\begin{bmatrix}
\omega_p & &  \\
& \omega_q & \\
& & \omega_r \\
\end{bmatrix} 
\begin{bmatrix}
p_{c} - p \\
q_{c} - q \\
r_{c} - r \\
\end{bmatrix} 
\end{equation}
where $\omega_p, \omega_q, \omega_r$ are control gains for $p$, $q$, and $r$ channels, respectively. Consequently, if $\dot{\phi} = p$, $\dot{\theta} = q$, and $\dot{\psi} = r$ are assumed while deriving the outer-loop control law for the sake of simplicity, the following expression is obtained as given in equation~\eqref{eq:eq14}.

\begin{equation} \label{eq:eq14}
\begin{bmatrix}
p \\
q \\
r
\end{bmatrix}_c
= 
\begin{bmatrix}
\omega_\phi & &  \\
& \omega_\theta & \\
& & \omega_\psi \\
\end{bmatrix} 
\begin{bmatrix}
\phi_{c} - \phi \\
\theta_{c} - \theta \\
\psi_{c} - \psi \\
\end{bmatrix} 
\end{equation}
in which $\omega_\phi, \omega_\theta, \omega_\psi$ are control gains for $\phi$, $\theta$, and $\psi$ channels, respectively. Deriving the control law in this manner is preferred for creating an ease in the derivation of the reconfiguration law, which will be introduced in the proceeding sections.

\subsection{Control allocation design}
\label{ca}

In order to capture the highly nonlinear aerodynamic characteristics of the aircraft, the incremental nonlinear control allocation method is preferred as the control allocation scheme. Therefore, the demanded control moment coefficients vector is incrementally converted into the corresponding control surface deflections. The formulation of the incremental nonlinear control allocation in the general sense is given in equation~\eqref{eq:eqINCA}.

\begin{equation} \label{eq:eqINCA}
\Phi(x_0, u_0) \Delta \bm{u} = \Delta \bm{\tau_c}
\end{equation}
where $\Phi(x_0, u_0) \in \mathbb{R}^{3 \times n}$ Jacobian matrix at that instant, which $n$ is the number of control effectors. The Jacobian matrix also refers to as the control effectivity matrix. Finally, $\Delta \bm{\tau}_c$ is the difference between the demanded control moment coefficient vector derived in the flight control module and the current moment coefficient. Since the aircraft is over-actuated, the resulting control effectivity matrix is not square; therefore, Moore-Penrose pseudo-inverse is utilized in the weighted form given in equation~\eqref{eq:eqWPI}.

\begin{equation} \label{eq:eqWPI}
    \begin{aligned}
        \Phi^{\ddag} = W \Phi^T (\Phi W \Phi^T)^{-1}
    \end{aligned}
\end{equation}
where $\Phi^{\ddag}$ is the pseudo-inverse of the control effectivity matrix, and $W \in \mathbb{R}^{3 \times 3}$ is the diagonal positive-definite weight matrix. Consequently, the required control surface deflections are calculated incrementally as given in equation~\eqref{eq:eqWPI2}.

\begin{equation} \label{eq:eqWPI2}
    \begin{aligned}
    \Delta \bm{u} = \Phi^{\ddag} \Delta \bm{\tau}_c
    \end{aligned}
\end{equation}

\begin{figure*}[hbt!]
\centering
\includegraphics[width=\textwidth,height=3.75cm]{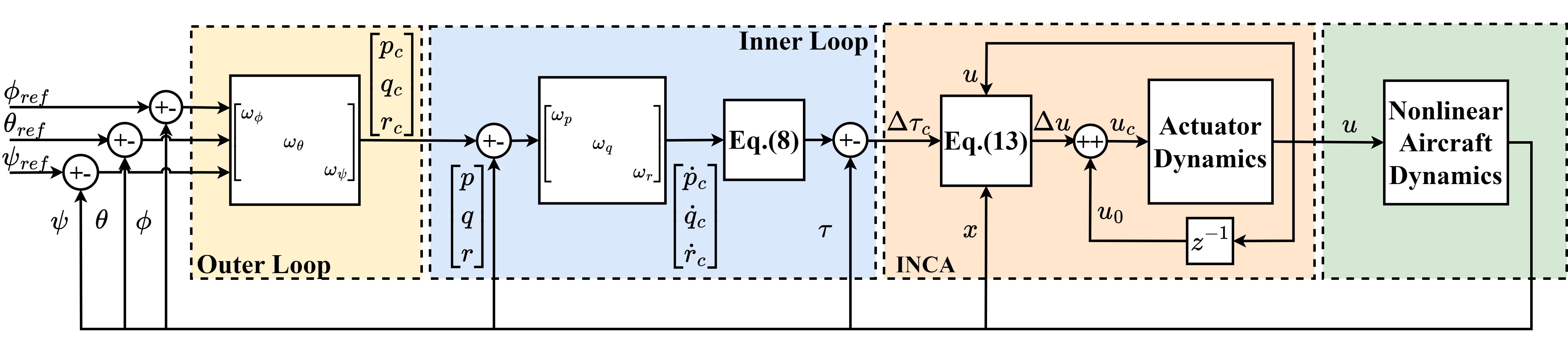}
\caption{Entire flight control architecture: nonlinear dynamic inversion based control law derivation and weighted pseudo-inverse based incremental nonlinear control allocation application to the nonlinear flight dynamics.}
\label{fig:fca}
\end{figure*}

At the end of the process, last actuator demand should be added to the calculated incremental actuator demand, i.e. $\Delta \bm{u} = \bm{u} - \bm{u}_0$. For the sake of clarity, entire flight control architecture is given in diagram represented in Fig.~\ref{fig:fca}.

As expected, the corresponding state feedback are supplied from the flight dynamics model. Also, as a note, $\phi_c, \theta_c, \psi_c$ and $p_c, q_c, r_c$ commands are passed through a second-order and first-order command filters, respectively.

\section{Reconfigurable Fault-tolerant Flight Control}
\label{rftfc}

The reconfiguration of the flight control architecture depends on two notions: incremental attainable acceleration set and bandwidth ratio, which are used for inner-loop redesign and outer-loop redesign respectively. Moreover, the processes introduced in the proceeding sections are activated only in the event of a fault diagnosis; otherwise, the aircraft operates with the preset flight control architecture.

\subsection{Inner-loop Reconfiguration}
\label{ilr}

\begin{figure}[hbt!]
\centering
\includegraphics[width=3.3in]{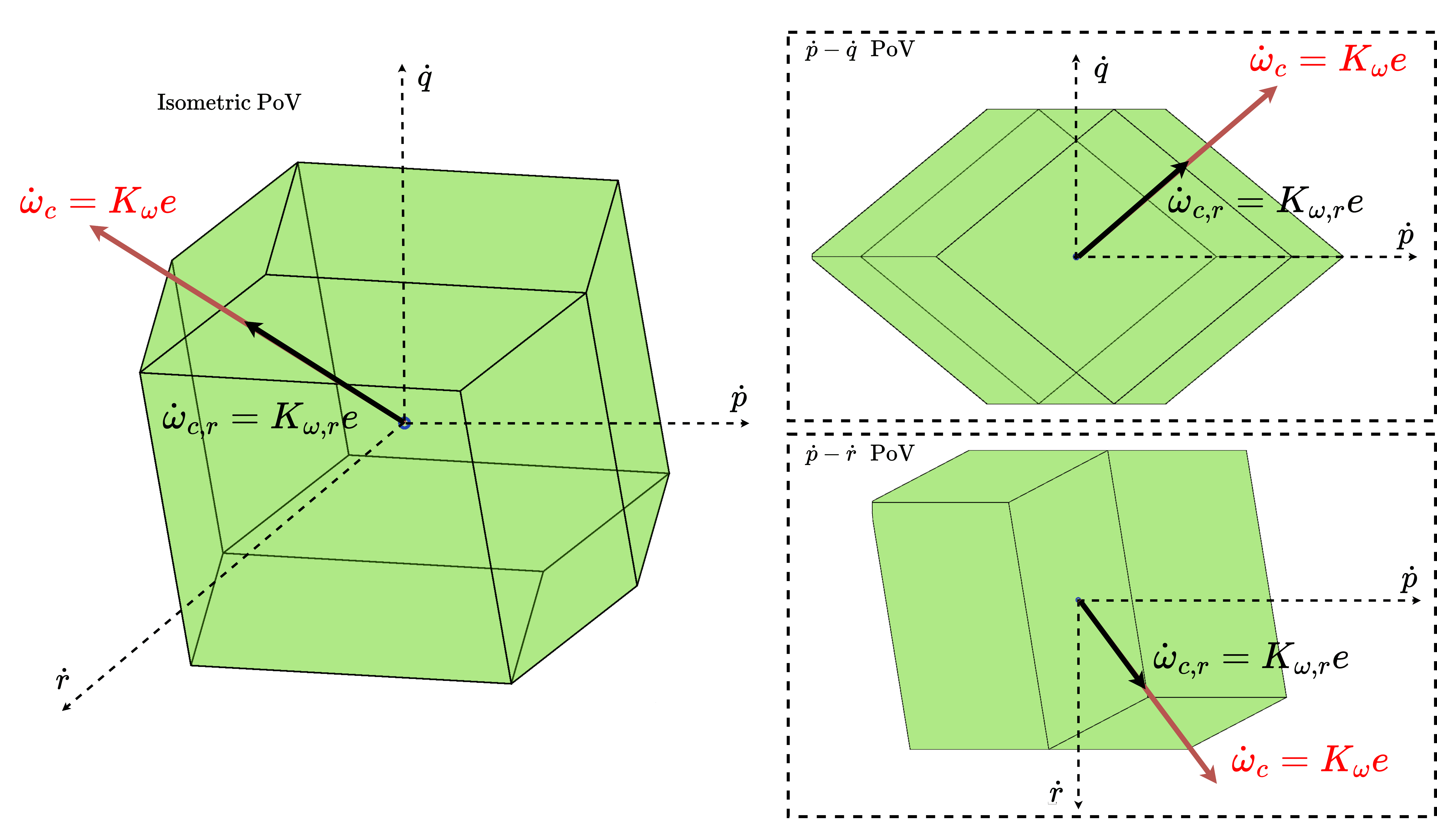}
\caption{Background for the reconfiguration strategy: A sample IAAS at an arbitrary time step including commanded acceleration vector, $\dot{\omega}_c$, which violates the boundary. Also, $\dot{\omega}_{c,r}$ is the reconfigured acceleration command by reconfiguring $K_\omega$ to $K_{\omega,r}$.}
\label{fig:aasRepresentative}
\end{figure}

The effect of the inner-loop on the aircraft is stabilizing it, meaning that the inner-loop operates as a stability augmentation system. Therefore, the controllability must be ensured in the inner-loop while reconfiguring the corresponding gains. At this point, as a useful tool, the incremental attainable aceleration set is benefited to create an adaptiveness of the inner-loop for fault accommodation. Prior to introducing the reconfiguration strategy, a representative illustration of the general framework is depicted in Fig.~\ref{fig:aasRepresentative}.

Obviously, it is possible to instantaneously stretch or shorten the desired angular acceleration vector by comparing them to the incremental attainable acceleration set at the corresponding time step. As a consequence, the desired angular acceleration vectors can be guaranteed to be bounded by the attainable set. In essence, this ensures controllability, as the boundaries of the incremental attainable acceleration set determine the maximum or minimum reachable accelerations without compromising controllability. However, there are two presumable issues during the recalculation process: the chattering due to the highly nonlinear aerodynamic characteristics and restricting the gain alteration within a specific interval. Consequently, the following reconfiguration strategy is established as given in equation~\eqref{eq:ilr1} by addressing the preceding issues.

\begin{equation} \label{eq:ilr1}
\begin{split}
\underbrace{
\begin{bmatrix}
\omega_{p_r} & & \\
& \omega_{q_r} & \\
&  & \omega_{r_r} \\
\end{bmatrix}}
_{\substack{\text{reconfigured}\\\text{gain, $K_{\omega,r}$}}}
= \cdots \quad \quad \quad \quad \quad \quad \quad \quad \quad \quad \quad \quad \quad \quad \quad \\
\underbrace{
\begin{bmatrix}
\sigma_p & &  \\
& \sigma_q & \\
& & \sigma_r
\end{bmatrix}}
_{\substack{\text{tuning}\\\text{functions, $\sigma_\omega$}}}
\underbrace{
\begin{bmatrix}
\dot{p}_{\Delta \mathbb{A_A}} & & \\
& \dot{q}_{\Delta \mathbb{A_A}} & \\
& & \dot{r}_{\Delta \mathbb{A_A}} \\
\end{bmatrix}}
_{\substack{\text{capability limits}\\\text{for each channel, $\Delta \mathbb{A_A}_\omega$}}}
\oslash
\underbrace{
\begin{bmatrix}
sgn(e_p)(|e_p| + \epsilon) \\
sgn(e_q)(|e_q| + \epsilon) \\
sgn(e_r)(|e_r| + \epsilon)
\end{bmatrix}}
_{\substack{\text{error for each channel, $\bm{e}$}}}
\end{split}
\end{equation}
where $\omega_{p_r}, \omega_{q_r}, \omega_{r_r}$ are the reconfigured gains of the angular velocities and $\epsilon \in \mathbb{R}_{>0}$ is small positive number to prevent the zero division. Additionally, $\dot{p}_{\Delta \mathbb{A_A}}$, $\dot{q}_{\Delta \mathbb{A_A}}$, $\dot{r}_{\Delta \mathbb{A_A}} \in \mathbb{R}$ are the local extremum magnitudes of each angular accelerations over the incremental attainable acceleration set, along the direction of the commanded acceleration vector. Specifically, if the error for the corresponding control channel is positive, the extremum value is positive; otherwise, vice versa holds as illustrated in Fig.~\ref{fig:aasRepresentative}. Finally, one of the unique aspects of the formulation is the utilization of tuning functions: $\sigma_p, \sigma_q$, and $\sigma_r$ which will be discussed in the following section.

\subsubsection{Tuning function design}
\label{tfd}

\begin{figure}[hbt!]
\centering
\includegraphics[width=3.5in]{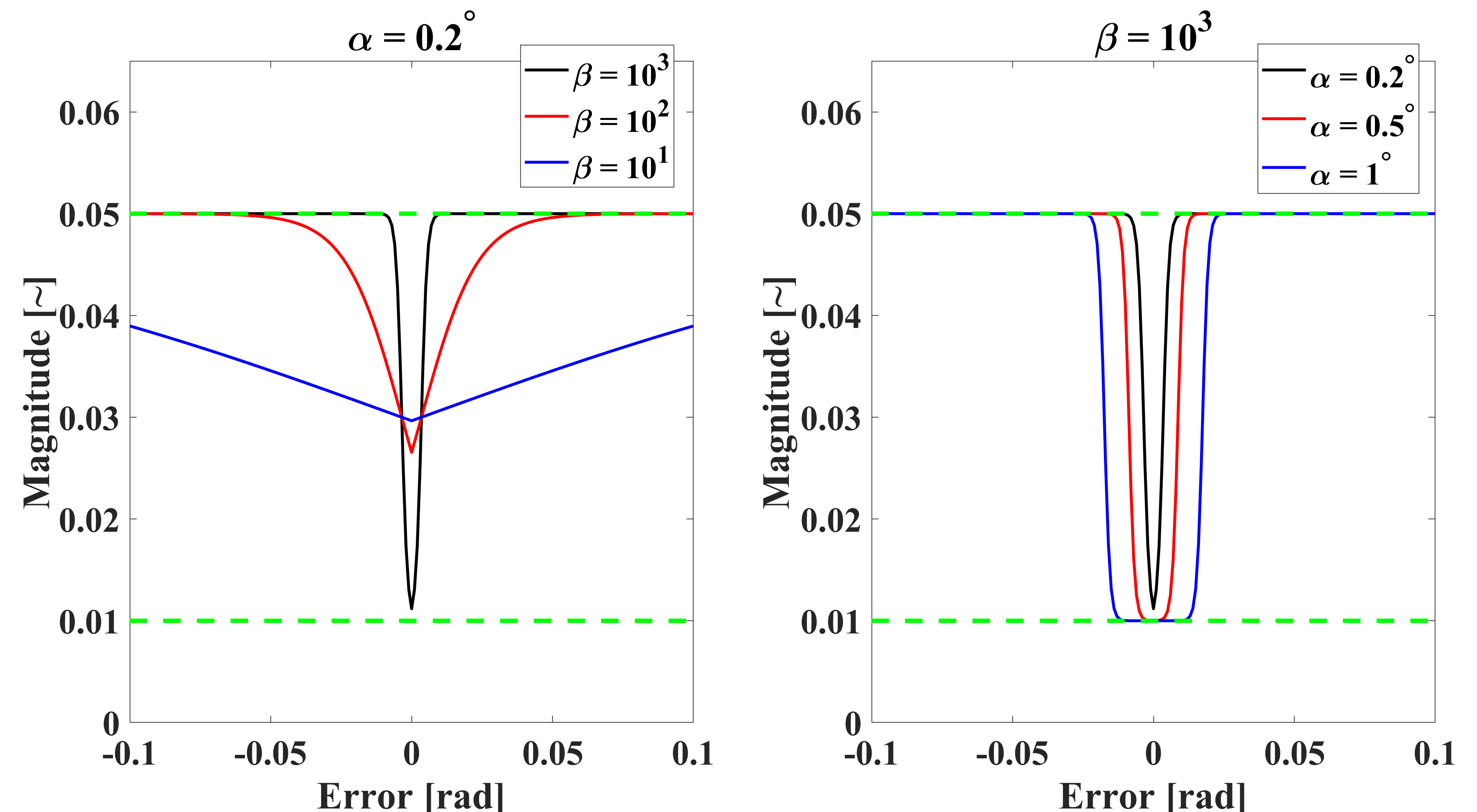}
\caption{Tuning function behaviour with the change of hyper-parameters.}
\label{fig:tuningFunction}
\end{figure}

In order to overcome the previously stated presumable issues, a sigmoid-like tuning function is designed, and indeed, tuning function is uniquely and separately designed for each inner-loop control channel. The formulation is given in equation~\eqref{eq:sigmoidlike}.

\begin{equation}\label{eq:sigmoidlike}
    \sigma_{\omega} = \sigma_{\omega_l} + \dfrac{\sigma_{\omega_u} - \sigma_{\omega_l}}{1 + e^{-\beta(|e_{\omega}| - \alpha)}}
\end{equation}
where $\sigma_{{\omega}_l}$ and $\sigma_{{\omega}_u}$ are the lower and upper limits of the tuning function for channel $\omega$, respectively. Also, $\alpha \in \mathbb{R}_{>0}$ and $\beta \in \mathbb{R}_{>0}$ are hyper-parameters to be tuned. The variation of the tuning function with the hyper-parameters of $\alpha$ and $\beta$ is demonstrated in Fig.~\ref{fig:tuningFunction}.

The purpose of the tuning function is to compensate for the adaptive gain's behavior concerning the error level. Therefore, it is designed as a function of the error in the corresponding channel. For example, when the error level is relatively low, the related gain tends to increase excessively due to the division by error within the attainable set. However, at that moment, the tuning function tends to decrease. As a result, the gain is prevented from increasing excessively.

On the other hand, due to the nonlinear characteristics of aerodynamics, the attainable set may exhibit chatter during unsteady flight. Therefore, the tuning function is also responsible for ensuring that the gain alteration is as smooth as possible. It's important to note that the tuning function uses hyper-parameters that need to be adjusted by the user. To achieve satisfactory performance in terms of both tracking and stability, these hyper-parameters should be carefully tuned, e.g. as $\alpha$ decreases, the reduction of the tuning function magnitude takes place closer to zero error, whilst as $\beta$ decreases the gain gets more sluggish, or in other words smoother.

\subsubsection{Inner-loop reconfiguration stability analysis}

Since the tracking error stabilization is required to ensure the maneuver stability, the Lyapunov candidate is constructed based on the tracking error, as presented in Eq.~\eqref{lyapunovCandidate}.

\begin{equation} \label{lyapunovCandidate}
    V(\bm{e}) = \frac{1}{2}\bm{e}^T\bm{e}
\end{equation}
where $\bm{e}(t) = \bm{\omega}_{ref}(t) - \bm{\omega}(t)$, the difference between the reference signal and obtained state. Obviously, $\bm{e}^* = \bm{0}_n$, $V(\bm{e}^*) = 0$, and $V(\bm{e}) \in \mathbb{R}_{>0}, \forall \bm{e} \in \mathbb{R}^n-\{0\}$. Simply, the time derivative of the Lyapunov candidate is derived as in Eq.~\eqref{timeDerivativeLyapunovCandidate}.

\begin{equation} \label{timeDerivativeLyapunovCandidate}
    \frac{dV(\bm{e})}{dt} = \frac{\partial V}{\partial \bm{e}} \frac{\partial \bm{e}}{\partial t} = \bm{e}^T\dot{\bm{e}}
\end{equation}

Since $\dot{\bm{e}} = \dot{\bm{\omega}}_{ref} - \dot{\bm{\omega}}$, Eq.~\eqref{timeDerivativeLyapunovCandidate} is rewritten in Eq.~\eqref{timeDerivativeLyapunovCandidate2}.

\begin{equation} \label{timeDerivativeLyapunovCandidate2}
    \dot{V} = \bm{e}^T\dot{\bm{\omega}}_{ref} - \bm{e}^T[f(\bm{\omega}) + g(\bm{\omega})\bm{u}]
\end{equation}

\begin{assumption}
    The reference signal is known, $C^1$ class function.
\end{assumption}

Since the control input of $\bm{u}$ is derived using the nonlinear dynamic inversion in equation~\eqref{eq:eq12}, i.e. $\bm{u} = \bm{g}(\bm{\omega})^{-1}[\bm{\Dot{\omega}}_c - \bm{f}(\bm{\omega})]$, the equation~\eqref{timeDerivativeLyapunovCandidate2} turns into equation~\eqref{timeDerivativeLyapunovCandidate3}.

\begin{equation} \label{timeDerivativeLyapunovCandidate3}
    \dot{V} = \bm{e}^T\dot{\bm{\omega}}_{ref} - \bm{e}^T \Dot{\bm{\omega}}_c < 0 
\end{equation}

Recall that $K_{\omega,r} = \sigma_\omega \Delta \mathbb{A_A}_\omega \oslash \bm{e}$ from equation~\eqref{eq:ilr1}. Therefore, the following expression is obtained in equation~\eqref{timeDerivativeLyapunovCandidate4}.

\begin{equation} \label{timeDerivativeLyapunovCandidate4}
    \dot{V} = \bm{e}^T\dot{\bm{\omega}}_{ref} - \bm{e}^T [\sigma_\omega \Delta \mathbb{A_A}_\omega \oslash \bm{e}] \bm{e} < 0 
\end{equation}

For the asymptotically stable error dynamics, the following inequality in equation~\eqref{timeDerivativeLyapunovCandidate5} must be satisfied.

\begin{equation} \label{timeDerivativeLyapunovCandidate5}
    \bm{e}^T\dot{\bm{\omega}}_{ref} < \bm{e}^T \sigma_\omega \Delta \mathbb{A_A}_\omega \bm{1}_3  
\end{equation}

Consequently, the following relation in equation~\eqref{stabilityPiecewise} concludes the stability requirements.

\begin{equation} \label{stabilityPiecewise}
    \lvert \dot{\bm{\omega}}_{ref,i} \rvert < \lvert \sigma_{\omega,i} \Delta \mathbb{A_A}_{\omega,i} \rvert \Rightarrow \Dot{V} < 0 \\
\end{equation}
where $i = {1,2,3}$ denotes the $p$, $q$, and $r$, respectively.

\begin{corollary}
    The relation in equation~\eqref{stabilityPiecewise} implies that there exist such a tuning function of $\sigma_\omega$ that establishes a balance between the desired input and capabilities of the aircraft. Furthermore, since the commanded acceleration vector must be bounded by the IAAS, the maximum value of the tuning function must be less than 1.
\end{corollary}

\subsection{Outer-loop reconfiguration}
\label{olr}

The outer-loop serves as the control augmentation system, while the inner-loop functions as the stability augmentation system. Consequently, once the inner-loop achieves its stabilization objective, the main goal of outer-loop reconfiguration should be to ensure overall closed-loop stability and achieve good tracking performance. To satisfy these expectations, reconfiguring the outer-loop gains through making them dependent on the inner-loop gains is interiorized.  

\begin{figure}[hbt!]
\centering
\includegraphics[width=3.25in]{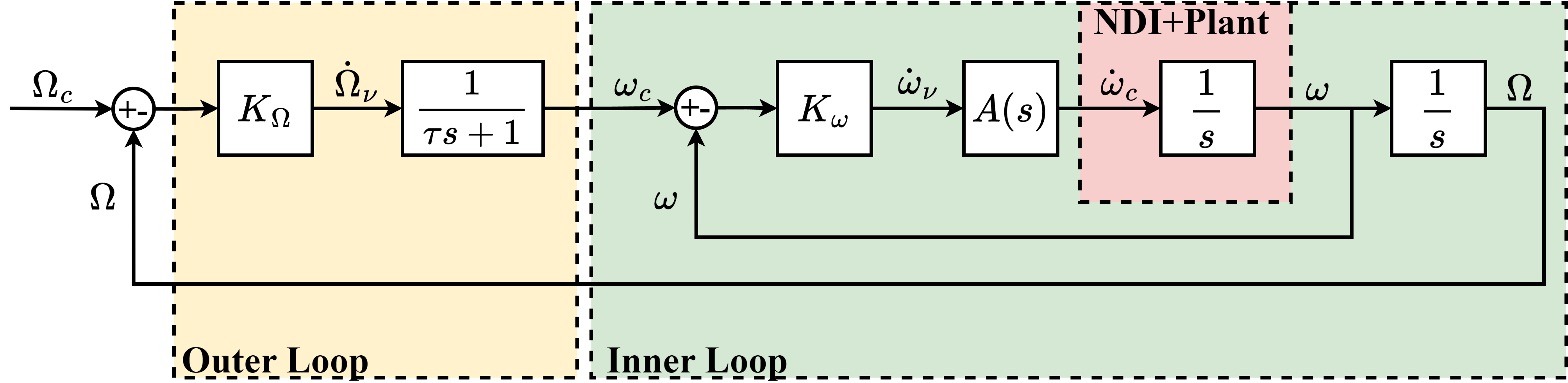}
\caption{The representation of the nonlinear closed-loop system in the linearized form due to the nonlinear dynamic inversion: $\Dot{\Omega}_\nu$ denotes the commanded virtual input, i.e. $\omega_c$, while $\Dot{\omega}_\nu$ also denotes the virtual input.}
\label{fig:olr1}
\end{figure}

The bandwidth concept is highly suitable for achieving each of the aforementioned goals. This is because keeping a certain and safe level of the bandwidth ratio between the inner and outer loops can ensure overall closed-loop stability. Additionally, bandwidth can be employed to adjust track performance, as it is well-known that bandwidth is inversely proportional to the damping ratio of the system. Consequently, it becomes possible to derive an analytical and explicit expression for the outer-loop gains in relation to the inner-loop gains. However, prior to delving into the derivation, let introduce the equivalent linear representation of the overall system as shown in Fig.~\ref{fig:olr1}.

in which $K_\Omega$ are the outer-loops gains, $K_{\omega}$ are the inner-loop gains, $A(s)$ is the actuator dynamics, $A(s) = 1/(\tau_a s + 1)$; $\tau$ and $\tau_{a}$ are the command filter and actuator time constants respectively. Since the nonlinear dynamic inversion is employed in the inner-loop control, the aircraft dynamics are transformed to a single integrator system; however, the actuator dynamics and command filters still appear. Moreover, since $\dot{\phi} = p$, $\dot{\theta} = q$, and $\dot{\psi} = r$ are assumed, the representation provided above remains valid for investigating. Subsequently, let derive the transfer functions for both inner-loop and overall system. The inner loop transfer function can be easily obtained since it is second-order simple system as given in equation~\eqref{eq:iltf} in a general sense.

\begin{equation}\label{eq:iltf}
    \dfrac{\omega}{\omega_{c}} = \dfrac{K_{\omega}}{\tau_a s^2 + s + K_{\omega}}
\end{equation}
where the natural frequency is, $\omega_n = \sqrt{K_{\omega}/\tau_a}$, and the damping ratio is, $\zeta = 1/\sqrt{4 K_{\omega} \tau_a}$. The bandwidth of a conventional second-order system is given in equation~\eqref{eq:bw1}.

\begin{equation}\label{eq:bw1}
    \nu = \omega_n \sqrt{1 - 2\zeta^2 + \sqrt{2 - 4\zeta^2 + 4\zeta^4}}
\end{equation}

If the obtained natural frequency and damping ratio are plugged in equation~\eqref{eq:bw1}, the following expression for the inner-loop bandwidth is obtained as given in equation~\eqref{eq:bw2}.

\begin{equation}\label{eq:bw2}
\begin{split}
    \nu_{\omega} = \sqrt{\dfrac{K_{\omega}}{\tau_a}}\sqrt{\sqrt{\dfrac{1}{4 K_{\omega}^2 \tau_{a}^2} - \dfrac{1}{K_{\omega} \tau_a} + 2} - \dfrac{1}{2 K_{\omega} \tau_a} + 1}
\end{split}
\end{equation}

Finally, the bandwidth of the inner-loop is expressed in terms of actuator time constant and inner-loop gain of the corresponding channel. At this point, let introduce the overall system transfer function to derive its bandwidth. The overall system closed-loop transfer function is given in equation~\eqref{eq:ostf}.

\begin{equation}\label{eq:ostf}
    \dfrac{\Omega}{\Omega_{c}} =
    \dfrac{K_{\Omega} K_{\omega}}{\tau_a \tau s^4 + (\tau_a + \tau) s^3 + (1 + K_{\omega} \tau) s^2 + K_{\omega} s + K_\Omega K_{\omega}}
\end{equation}

Recall that the bandwidth frequency is the point where the gain of the system is equal to the $1/\sqrt{2}$ of the gain at zero-frequency input, thus if the equation of $|G(j\nu_\Omega)| = \dfrac{1}{\sqrt{2}}|G(j0)|$ is set and $s = j\nu_\Omega$ and $s = j0$ are plugged in the equation~\eqref{eq:ostf}, the following final expression is obtained as given in equation~\eqref{eq:bw3}.

\begin{equation}\label{eq:bw3}
\begin{split}
    2 K_\Omega^2 K_{\omega}^2 - [\tau_a \tau \nu_{\Omega}^4 - \nu_{\Omega}^2 - K_{\omega} \tau \nu_{\Omega}^2 + K_\Omega K_{\omega}]^2 \cdots \\
    \cdots - [(-\tau_a - \tau)\nu_{\Omega}^3 + K_{\omega} \nu_{\Omega}]^2 = 0 
\end{split}
\end{equation}
where $\nu_{\Omega}$ is the overall system bandwidth for the corresponding channel.

If the equation~\eqref{eq:bw3} is solved for $\nu_{\Omega}$, the bandwidth can be calculated; however, if the equation is solved for $K_\Omega$ in order to obtain an explicit relationship between the inner and outer-loop gains, the following non-trivial lengthy expression is achieved as given in equation~\eqref{eq:bw4}.

\par\noindent\rule{\dimexpr(0.5\textwidth-0.5\columnsep-0.4pt)}{0.4pt}%
\rule{0.4pt}{6pt}

\begin{strip}
\begin{equation}\label{eq:bw4}
    \begin{split}
        K_\Omega = \dfrac{-\nu_{\Omega}^2 + \nu_{\Omega} \sqrt{2 K_{\omega}^2 \tau^2 \nu_{\Omega}^2 + K_{\omega}^2 - 4 K_{\omega} \tau_a \tau^2 \nu_{\Omega}^4 - 2 K_{\omega} \tau_a \nu_{\Omega}^2 + 2 K_{\omega} \tau \nu_{\Omega}^2 + 2 \tau_a^2 \tau^2 \nu_{\Omega}^6 + \tau_a^2 \nu_{\Omega}^4 - 2 \tau_a \tau \nu_{\Omega}^4 + \tau^2 \nu_{\Omega}^4 + 2\nu_{\Omega}^2}}{K_{\omega}} \cdots \\
        \cdots + \dfrac{\tau_a \tau \nu_{\Omega}^4 - K_{\omega} \tau \nu_{\Omega}^2}{K_{\omega}}
    \end{split}
\end{equation}
\end{strip}

\vspace{\belowdisplayskip}\hfill\rule[-6pt]{0.4pt}{6.4pt}%
\rule{\dimexpr(0.5\textwidth-0.5\columnsep-1pt)}{0.4pt}

As a result, an analytical update law of the outer-loop gains based on the inner-loop gains, actuator and command filter time constants, and overall closed-loop system bandwidth could be achieved. Therefore, if the overall closed-loop system bandwidth is specified while considering the bandwidth ratio of the inner-loop and overall system, calculating the outer-loop gain in-flight becomes a straightforward task.

\subsection{Reconfiguration Strategy}

Subsequent to diagnosing the actuator jam fault 0.1 seconds after the fault occurrence by a hypothetical fault detection and isolation module, the preset gains in the flight control unit are initialized to be reconfigured by considering the proposed methods. Principally, the inner-loop reconfiguration is achieved at each time step in an online manner by the incremental attainable acceleration set and a compensator tuning function in order to ensure the system stability to prevent a presumable loss of control due to the actuator fault. Also, note that the inner-loop bandwidth is calculated at each time step using equation~\eqref{eq:bw2}. 

Afterwards, in order to ensure the stability without sacrificing the agility, the bandwidth ratio is restricted to a level of 15, i.e. $\nu_{\omega}/\nu_\Omega =$ 15. As a consequence, the overall system's bandwidth is determined in that manner and thanks to the prior knowledge of the actuator and command filter time constants, the outer-loop gains are calculated by plugging these obtained parameters in equation~\eqref{eq:bw4}.

Finally, the inner-loop and outer-loop gains are reconfigured in-flight to guarantee the controllability, stability, and satisfactory track performance.

\section{Results}
\label{res}

\begin{figure}[hbt!]
\centering
\includegraphics[width=3.5in]{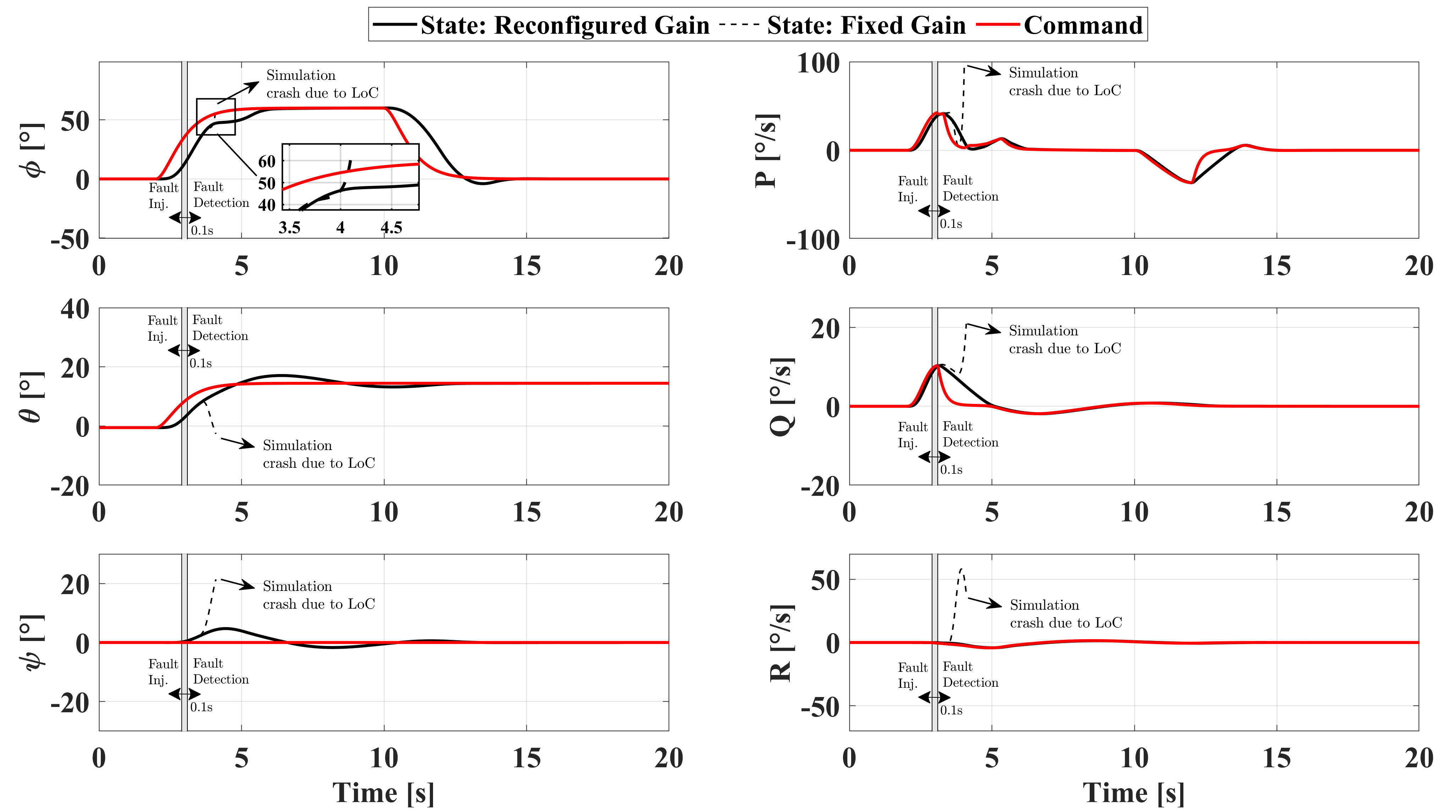}
\caption{State history of the test maneuver: reconfigured gain and fixed gain.}
\label{fig:man1_1}
\end{figure}

\begin{figure}[hbt!]
\centering
\includegraphics[width=3.5in]{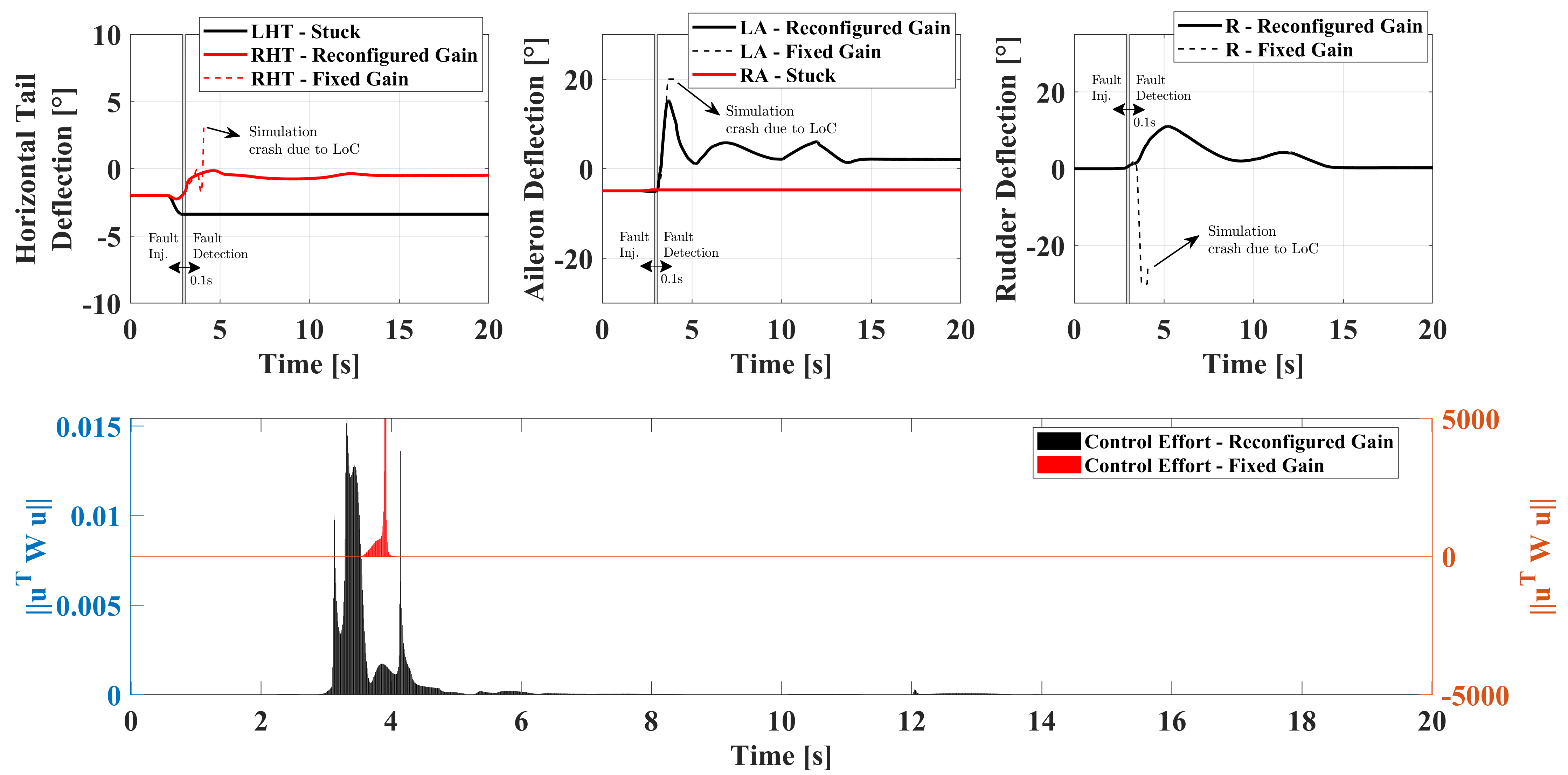}
\caption{Control deflection history and control effort of the test maneuver: reconfigured gain and fixed gain.}
\label{fig:man1_2}
\end{figure}

The efficacy of the proposed method is demonstrated through an excessive test case simulation, which encompasses both the right aileron and left horizontal tail simultaneous lock-in-place fault injection at the \nth{3} second of the simulation\footnote{For the simulation videos of more test cases, please visit: https://www.youtube.com/watch?v=7-TXah7Htbs}. However, the aircraft begins conducting a coupled commanded maneuver at \nth{2} second of the simulation, meaning that the lock-in-place occurs at a quite critical time instant. The corresponding flight simulation state trajectories and control surface deflection histories are illustrated in Fig.~\ref{fig:man1_1} and Fig.~\ref{fig:man1_2}.

\begin{figure}[hbt!]
\centering
\includegraphics[width=3.5in]{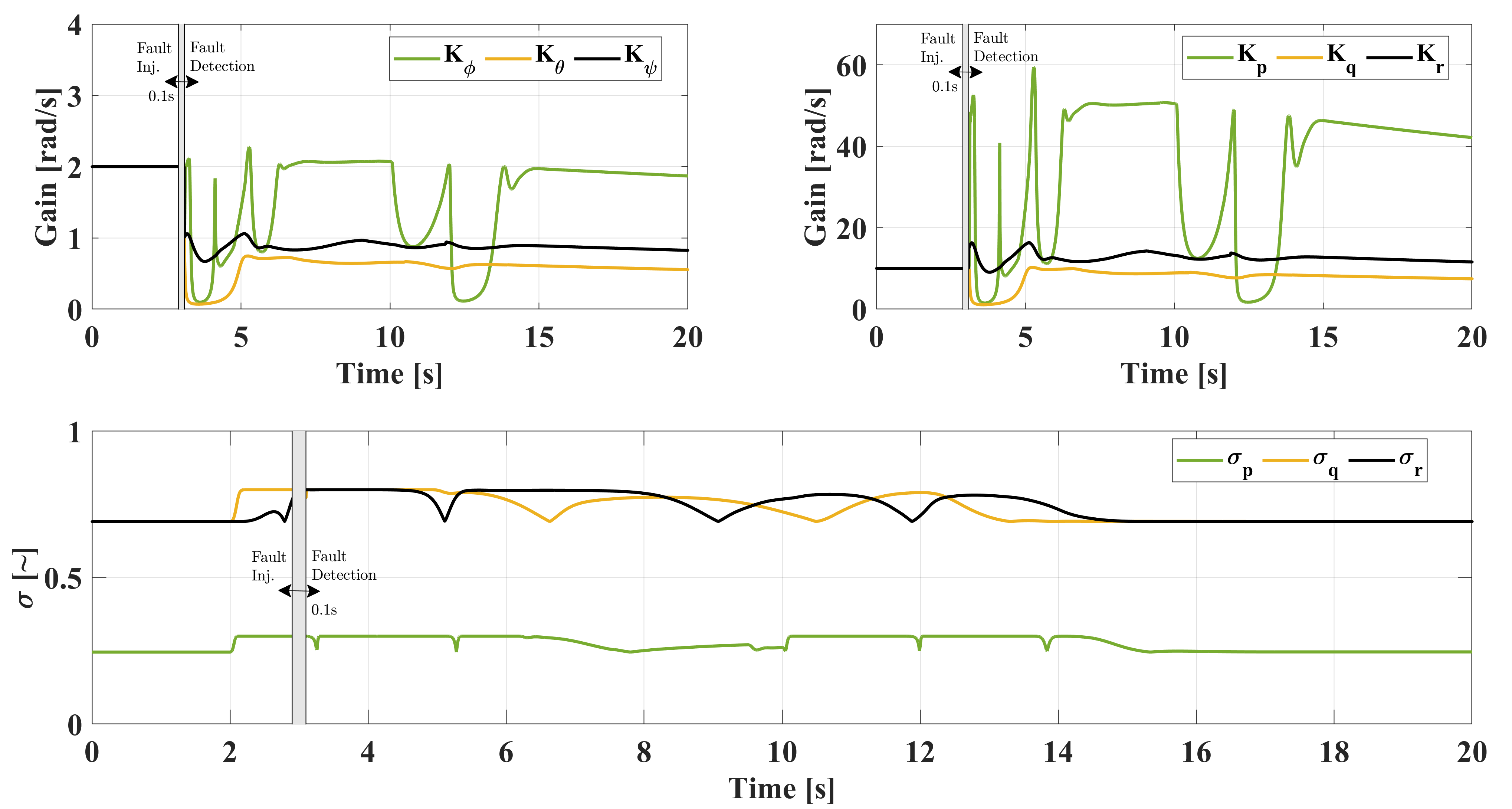}
\caption{Gains and tuning function history of the corresponding channels.}
\label{fig:man1_3}
\end{figure}

Obviously, being conservative by increasing the bandwidth ratio is a prudent step to ensure the system's stability in the event of a fault. The bandwidth ratio of the fixed-gain system is relatively low; therefore, the aircraft is more responsive. As a consequence, the aircraft with fixed-gain approach has a tendency to the loss of control. In contrast, the aircraft with reconfigured-gain approach is in a trade-off between robustness and agility, since it is bounded by the controllability, stability, and track performance metrics. Furthermore, the updated inner and outer-loop gain histories are presented in Fig.~\ref{fig:man1_3}.

\begin{figure}[hbt!]
  \centering
  \begin{tabular}[b]{c}
    \includegraphics[width=3.2in]{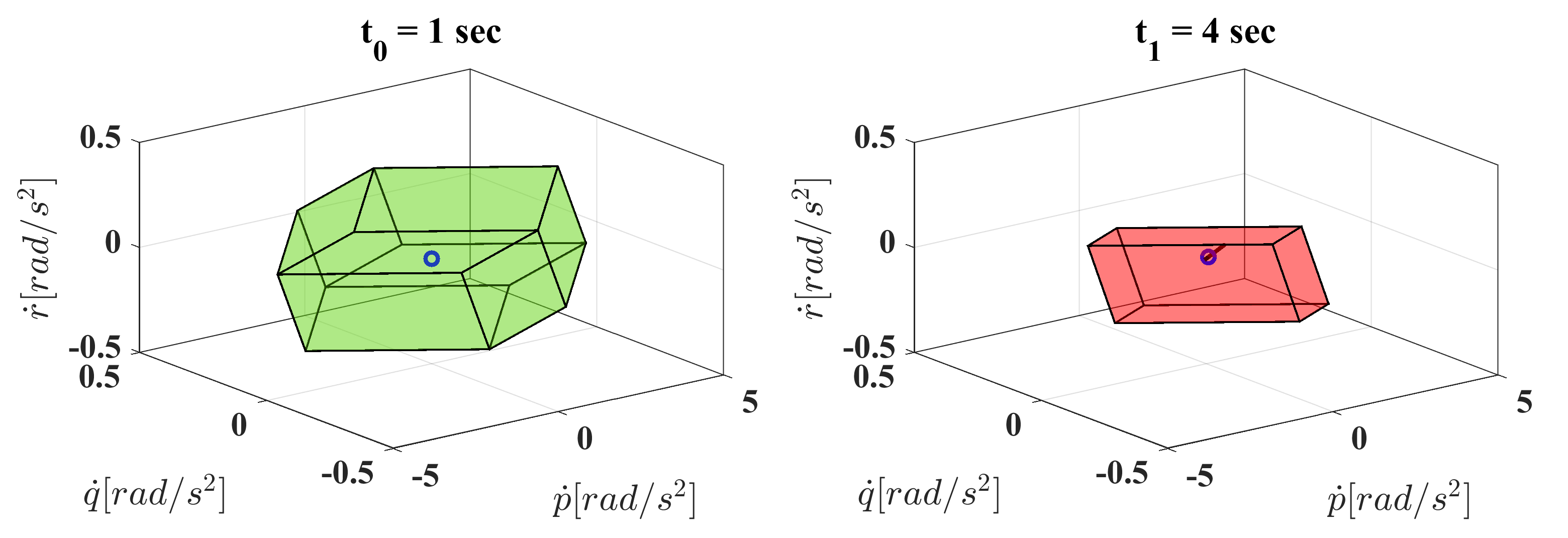} \\
    \small (a) Fixed gain during maneuver
  \end{tabular} 
  \begin{tabular}[b]{c}
    \includegraphics[width=3.2in]{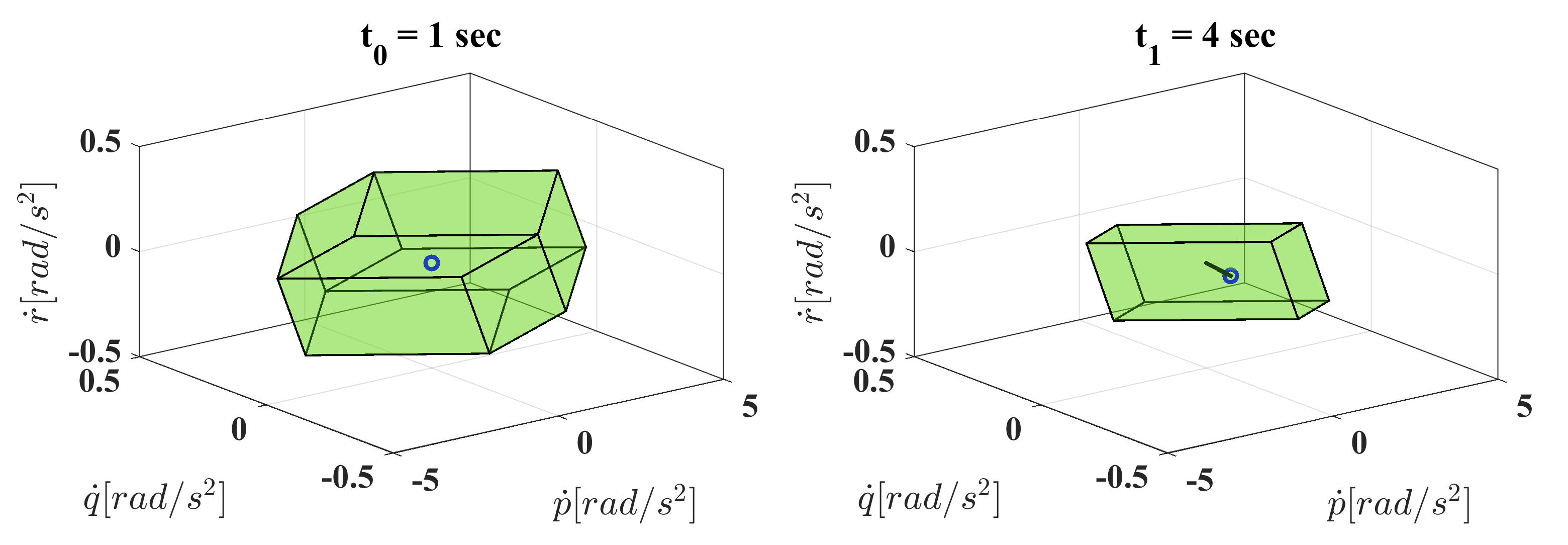} \\
    \small (b) Reconfigured gain during maneuver
  \end{tabular}  
   \vspace*{4mm}
  \caption{Attainable acceleration set history: (a) fixed gain approach, (b) reconfigured gain approach (\textbf{red:} if the control acceleration vector exceeds the IAAS; \textbf{green:} if the control acceleration vector is bounded by IAAS).}
  \label{fig:man1_4}
\end{figure}

\begin{figure}[hbt!]
\centering
\includegraphics[width=3.5in]{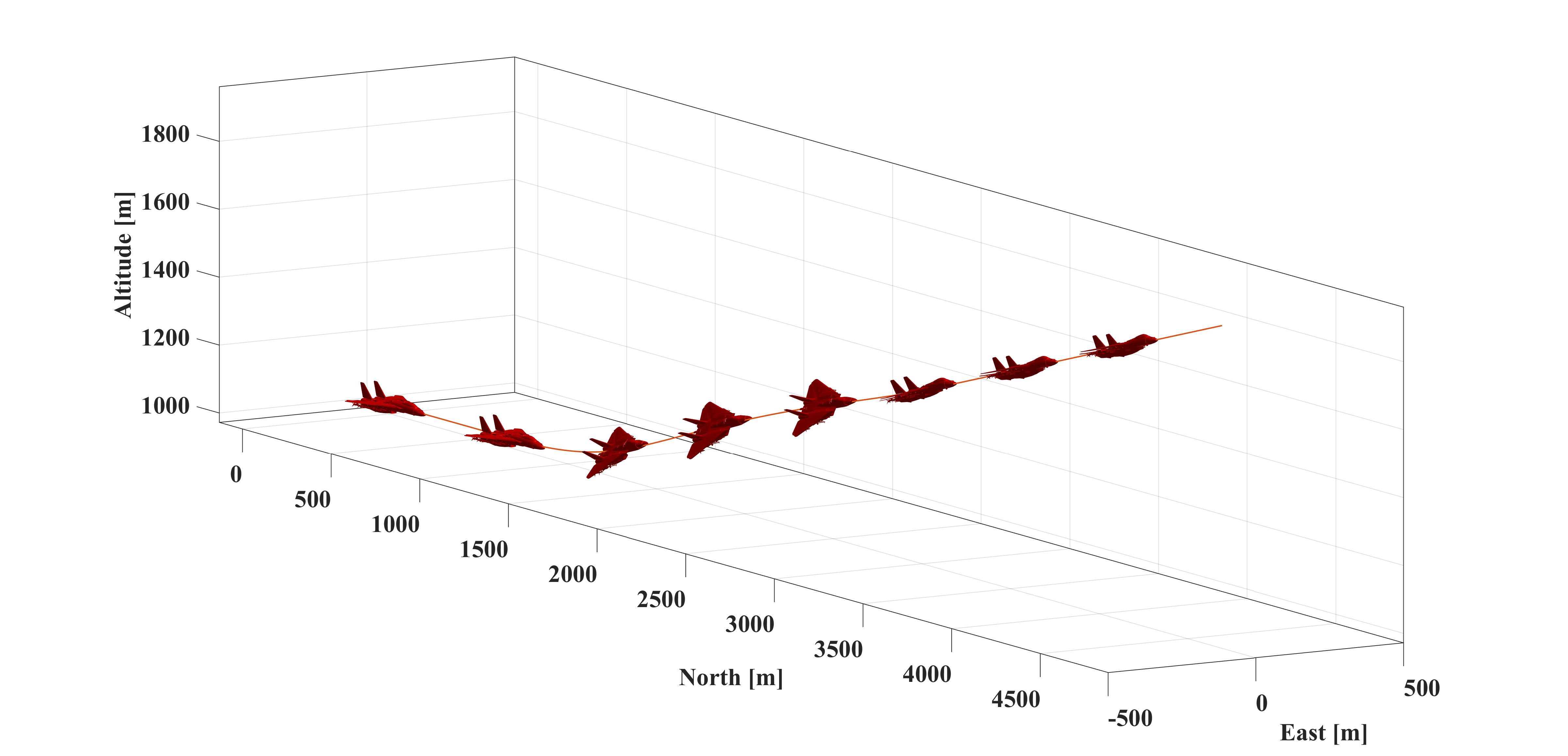}
\caption{3D trajectory of the maneuver.}
\label{fig:man1_5}
\end{figure}

Due to the ample volume of IAAS, it becomes possible to generate larger inner-loop gains. Consequently, the gains in the angular rate channels rise to an acceptable limit. In contrast, outer-loop gains typically decrease, often falling below their initial values to achieve the desired bandwidth ratio. It is noteworthy that the sigmoid-like tuning functions play a crucial role in adjusting the inner-loop gains, taking into account the boundaries defined by the IAAS, track performance expectations, and the need for smooth variations over time. Finally, the IAAS variations with time for both approaches are presented, afterwards the output trajectory of the flight is demonstrated in Fig.~\ref{fig:man1_4} and Fig.~\ref{fig:man1_5}, respectively.

Obviously, an unbounded control acceleration demand triggers a loss of control; nevertheless, designing the inner-loop gains in such manner that the generated control acceleration vector remains bounded by IAAS of the aircraft can mitigate the risk of the loss of control subsequent to a fault. Also, noticeably, the actuator fault causes IAAS to shrink and results in disappearance of several vertices, i.e. controllable bounds are seriously deteriorated.

Consequently, it is observed that the proposed method is highly satisfactory in terms of both track performance and stability.

\section{Conclusion}
\label{conc}

Throughout this study, a unique active fault-tolerant flight control method is thoroughly examined, designed to ensure the stability of the aircraft even under demanding mission conditions. The inner-loop gains are updated using the introduced notion, i.e. incremental attainable acceleration set, to ensure that the requested control acceleration vectors remain bounded. Furthermore, a more conservative formulation is derived for updating the outer-loop gains based on the inner-loop gain, actuator and command filter time constants, and a predetermined bandwidth ratio \textemdash specifically, $\nu_{\omega}/\nu_\Omega =$ 15 in this study. Finally, the proposed architecture is assessed using a high-fidelity over-actuated F-16 nonlinear flight dynamics model running at 100Hz. The result has demonstrated that the proposed method is capable of stabilizing the aircraft even in the presence of extremely harsh fault and complex mission requirement. As an extension of this work, we envision applying the same approach to a real-world platform and conducting real-time experiments to further validate its effectiveness.

\section{Acknowledgement}
    The authors propose their gratitude to Dr. Mustafa Demir for his fruitful contributions to the study.

\bibliographystyle{IEEEtran}
\bibliography{sample}

\end{document}